\documentclass[utf8]{styles/FrontiersinHarvard}

\usepackage{url,hyperref,lineno,microtype,subcaption}
\usepackage[onehalfspacing]{setspace}
\usepackage{xcolor}

\def\keyFont{\fontsize{8}{11}\helveticabold }
\def\firstAuthorLast{Minhui Zou {et~al.}} 
\def\Authors{Minhui Zou\,$^{1,*}$, Nan Du\,$^{2,3}$ and Shahar Kvatinsky\,$^{1}$}
\def\Address{
    $^{1}$ Viterbi Faculty of Electrical and Computer Engineering, Technion – Israel Institute of Technology, Haifa, Israel 3200003 
    \\
    $^{2}$ Institute for Solid State Physics, Friedrich Schiller University Jena, Fürstengraben 1, 07743 Jena, Germany
    \\
    $^{3}$ Department of Quantum Detection, Leibniz Institute of Photonic Technology (IPHT), Albert-Einstein-Str. 9, 07745 Jena, Germany
    }
\def\corrAuthor{Minhui Zou}
\def\corrEmail{minhui@campus.technion.ac.il}

\begin{document}
\onecolumn
\firstpage{1}

\title[Security Techniques for Memristive Computing]{Review of Security Techniques for Memristor Computing Systems\footnote{This paper (doi: https://doi.org/10.3389/femat.2022.1010613) was published in Frontiers in Electronic Materials, section Semiconducting Materials and Devices under the CC-BY license (https://creativecommons.org/licenses/by/4.0/)}} 

\author[\firstAuthorLast ]{\Authors} 
\address{} 
\correspondance{} 

\extraAuth{}

\maketitle

\begin{abstract}
    Neural network (NN) algorithms have become the dominant tool in visual object recognition, natural language processing, and robotics. 
    To enhance the computational efficiency of these algorithms, in comparison to the traditional von Neuman computing architectures, researchers have been focusing on memristor computing systems.
    A major drawback when using memristor computing systems today is that, in the artificial intelligence (AI) era, well-trained NN models are intellectual property and, when loaded in the memristor computing systems, face theft threats, especially when running in edge devices. 
    An adversary may steal the well-trained NN models through advanced attacks such as learning attacks and side-channel analysis. 
    In this paper, we review different security techniques for protecting memristor computing systems. 
    Two threat models are described based on their assumptions regarding the adversary's capabilities: a black-box (BB) model and a white-box (WB) model. 
    We categorize the existing security techniques into five classes in the context of these threat models: thwarting learning attacks (BB), thwarting side-channel attacks (BB), NN model encryption (WB), NN weight transformation (WB), and fingerprint embedding (WB). 
    We also present a cross-comparison of the limitations of the security techniques.
    This paper could serve as an aid when designing secure memristor computing systems.

    \tiny
    \keyFont{ \section{Keywords:} neural network, memristor computing system, hardware security, theft threat, defense technique, neuromorphic} 
\end{abstract}

\section{Introduction}
\begin{figure}
        \begin{center}
        \includegraphics[width=17cm]{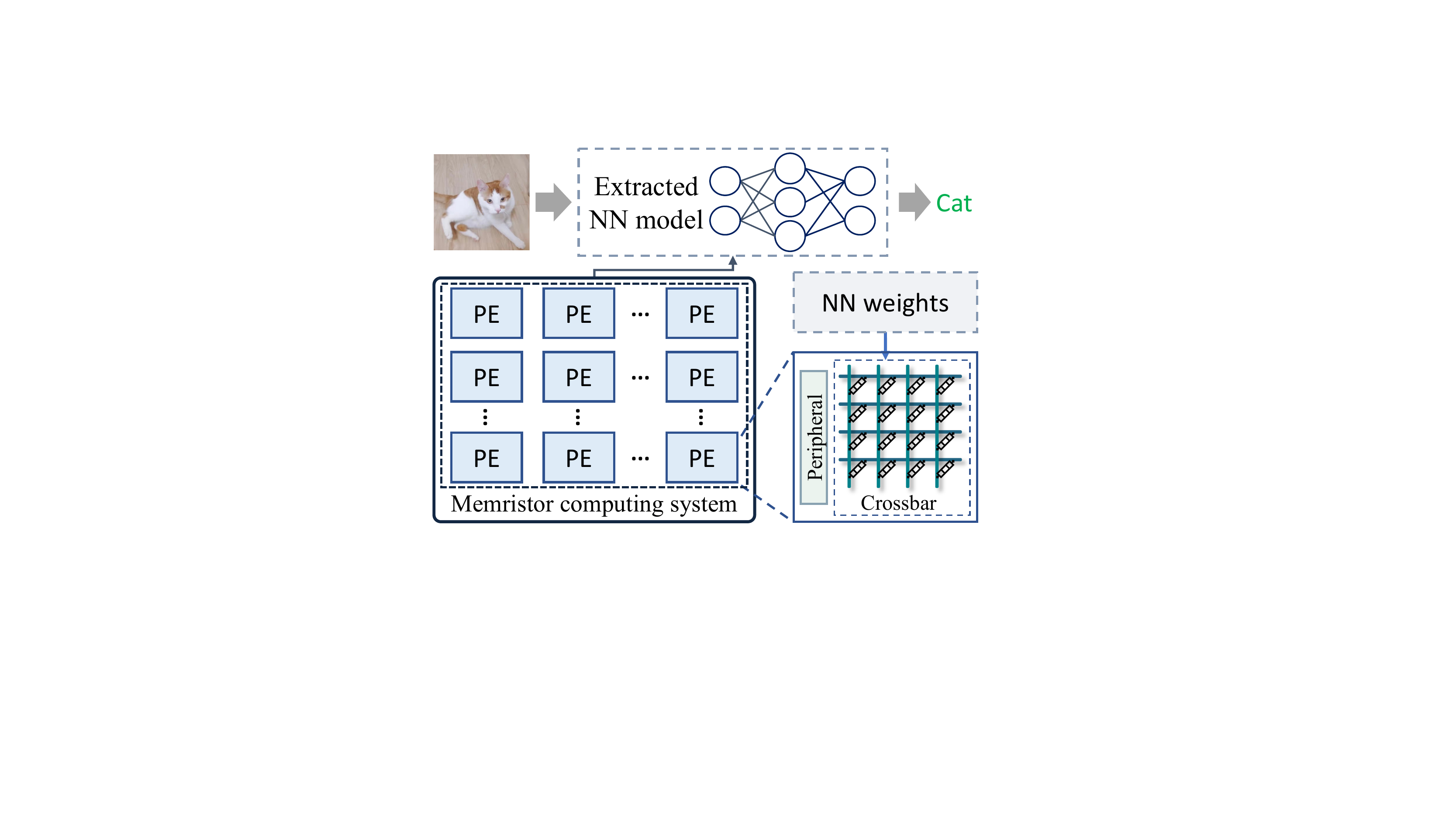}
        \end{center}
        \caption{Basic structure of memristor computing systems and theft attack models of loaded NN models: BB and WB} 
        \label{fig:basic_idea}
    \end{figure}

    Neural network (NN) algorithms have demonstrated great potential in certain areas such as visual object recognition, natural language processing, and robotics \cite{dong2021survey,karoly2020deep}.
    These algorithms involve a large number of vector-matrix multiplications (VMMs) and hence they are both data- and computing-intensive.
    Conventional computer architectures are designed by following the von Neumann model wherein the computation unit and memory are separate.
    When running on conventional computer architectures, NN algorithms require huge amounts of matrix data to be moved between the computational unit and memory, which demands a lot of time and energy.
    To overcome this obstacle, researchers and industry have turned to the emerging computing systems based on memristor devices.
    These systems could improve the efficiency of NN algorithms because they process the VMM operations directly in the memory so that data movement between the computing unit and memory is avoided.
    Moreover, the memristor devices could be structured in the form of a crossbar array. The input signals at the rows (wordlines, WLs) go through the memristor cells to accumulate at the columns (bitlines, BLs), which naturally analogizes VMM operations \cite{markovic2020physics}.
    The analogous VMM operations can be finished in a constant number of clock cycles, i.e., their time complexity is $O(1)$ \cite{li2015rram}, which further significantly improves the efficiency of the NN algorithms.
    As shown in Fig. \ref{fig:basic_idea}, a memristor computing system contains many processing elements (PE), with each PE comprising a memristor crossbar and peripheral circuits.
    The NN weights are mapped to the memristor crossbars.
    Thanks to the nonvolatile nature of memristor devices, the mapped NN weights will not be lost when the systems are rebooted and thus no remapping is needed.
    Advanced memristor computing systems exploit the intra- and inter- parallelism of memristor crossbars to boost the energy efficiency of different NN algorithms such as convolutional NN \cite{wen2020ckfo,wen2019memristor}, graph NN \cite{lyu2022efficient}, and spiking NN \cite{rathi2022exploring}.
    Various hardware-software co-optimization techniques have been proposed to push memristor computing systems to their energy efficiency limits.
    For example, \cite{wen2020ckfo,wen2019memristor} suggest pruning NN parameters with improved convolution algorithms to reduce the number of required memristors.
    These advantages of memristor computing systems make them promising for edge devices with constrained computational and energy budgets.
    
    Training an NN model is a computationally heavy task that demands a great deal of energy and financial resources.
    It may take days or even months \cite{strubell2019energy} and cost up to more than one million US dollars \cite{sharir2020cost} to output a well-trained NN model from scratch.
    Furthermore, large NN models involve a huge amount of training data.
    For example, GPT-3 was trained on 570 GB of filtered data \cite{brown2020language}.
    With NN models becoming larger and deeper, NN model training is becoming less advantageous for those who cannot afford the high training cost or the large training datasets.
    Additionally, if the training datasets are proprietary, there is always the risk that the trained NN models may leak confidential information contained in the datasets \cite{rajasekharan_scanet_2021}.
    Nowadays, well-trained NN models are deemed to be intellectual property and protecting them from being theft is imperative.

    NN models, loaded in memristor computing systems, as shown in Fig. \ref{fig:basic_idea}, may be vulnerable to theft threats such as learning attacks \cite{tramer_stealing_2016} and side-channel attacks \cite{hua2018reverse}. 
    Even worse, while the memristor devices’ nonvolatility might be a appealing feature, it facilitates data theft attacks.
    In the scenario of using memristor devices as main memory, data theft attacks have been widely considered as real threats \cite{young2015deuce,awad2016silent,awad2019triad,zuo2019supermem}.
    In memristor computing systems, the data persistence of the memristor devices may also expose the NN weights stored on the memristor crossbars to an adversary.
    Besides, \cite{huang_new_2020} claimed it is also possible to able to read the stored NN weights from the systems through micro-probing the peripheral circuits of memristor crossbars.
    All these threats represent serious security challenges to memristor computing systems.

    In this paper, we review different security techniques for protecting NN models for memristor computing systems.
    Previous works either focused on NN security techniques from the viewpoint of software \cite{oseni2021security} or security threats facing the memristor computing system hardware \cite{hu2022systematic}. 
    This review, in contrast, focuses on memristor computing hardware defense security techniques.
    The structure and major contributions of this paper are summarized as follows:

    \begin{itemize}
        \item We categorize the existing security techniques into five classes defined by the black-box (BB) and white-box (WB) threat models.
        \item We present a brief overview of the existing security techniques under the above categorization.
        \item We present a cross-comparison of the limitations of the various security techniques.
        \item We discuss the challenges of the existing countermeasures and suggest future research directions in this field.
    \end{itemize}

\section{Background}
    \subsection{Preliminaries}
        The most computationally heavy and time-consuming parts of NN algorithms are their convolution (Conv) layers and fully-connected (FC) layers.
        The main work of FC layers can be implemented directly with VMMs, described as: 
        \begin{equation} \label{eq:fc_operations}
            y_j = \sum_{i=1}^{m} x_{i} \cdot w_{i,j},
        \end{equation}
        where $x_{i}(i \in [1,m])$ is the input feature map, $y_j(j \in [1,n])$ is the output, and $w_{i,j}$ is the synapse weight.
        The main work of Conv layers is different but could also be transformed to be implemented with VMMs.
        $x_{i}$ and $y_j$, respectively, are still the input feature map and the output while each column of the weight matrix is a vector transformed from a filter kernel.
        To simplify our discussion, we assume the weights of the Conv layers are already transformed into matrices. 
        Thus, the weights of both the FC layers and the Conv layers are in the form of matrices.
    
        \begin{figure}[!ht]
            \begin{center}
            \includegraphics[width=8cm]{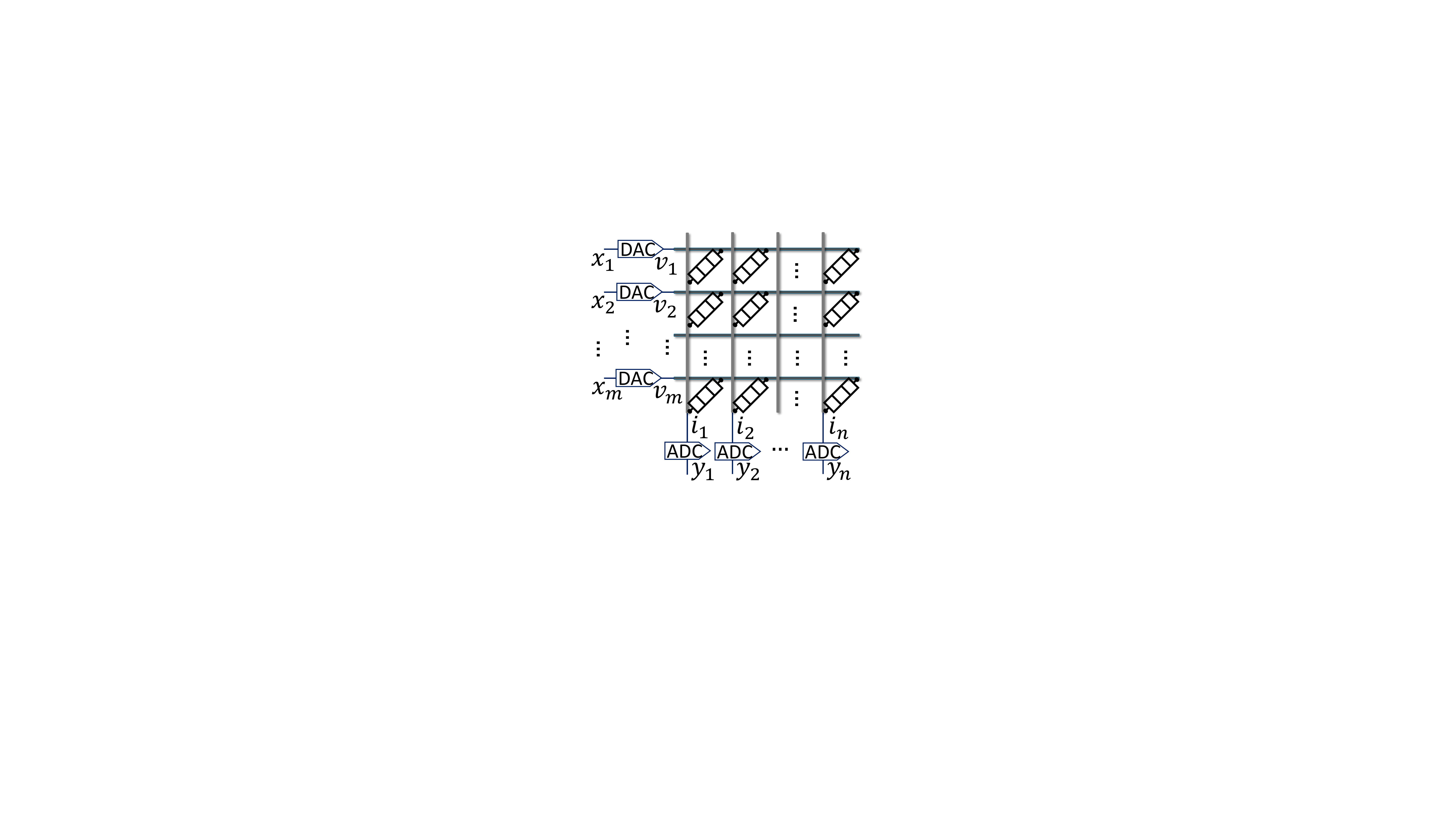}
            \end{center}
            \caption{A memristor crossbar executing analogous VMM by accumulating currents at its BLs: the accumulated current at the $j$th BL is $\sum_{i=1}^{m} v_{i} \cdot c_{i,j}$.} 
            \label{fig:vmm}
        \end{figure}
    
        In memristor computing systems, as shown in Fig. \ref{fig:vmm}, the input feature maps are transformed into voltages $(\textit{\textbf{v}})$ by using digital-to-analog converters (DACs) that are applied to the WLs of the memristor crossbars.
        The BLs of the memristor crossbars output the accumulated currents $(\textit{\textbf{i}})$.
        The output currents are then transformed by using analog-to-digital converters (ADCs).
        The analogous VMMs performed by a memristor crossbar are described as: 
        \begin{equation} \label{eq:analog_VMM}
            i_j = \sum_{i=1}^{m} v_{i} \cdot c_{i,j},
        \end{equation}
        where $c_{i,j}$ is the conductance of the cell at the crossbar's $i$th row and $j$th column.
        Memristor devices can be categorized as either analog \cite{wu2018methodology} or digital \cite{prakash2014demonstration}. 
        Analog memristor devices move gradually through a RESET process, which means an memristor device can be tuned from low resistance state (LRS) to high resistance state (HRS) continuously. 
        Consequently, an ideal analog memristor cell could be tuned into any arbitrary conductance state between LRS to HRS. 
        Digital memristor devices can only be tuned to limited discrete resistance states \cite{prakash2014demonstration}.
        Accordingly, multiple crossbars are used to represent high-precision weights \cite{zhu2019configurable,cai2019low,zhu2020mnsim}. 
        An NN weight can be positive or negative, but the conductance of memristor devices can only be positive. 
        To support negative weights, different mapping schemes were proposed. 
        Two popular mapping schemes are biasing the original weights to be non-negative (bias-based mapping) \cite{shafiee2016isaac,xue202015} and using the differential values of pairs of memristor devices to represent the original weights (differential mapping) \cite{chi2016prime,zhu2019configurable}.
        Besides, NN algorithms are typically processed layer by layer, and the outputs of one layer are used as the inputs of the next layer.
        Large NN layers are assigned to multiple PEs, and the partial sums are aggregated in a global buffer \cite{long2019ferroelectric,krishnan2021impact}.
        To maximize the processing parallelism of memristor crossbars, some memristor computing systems \cite{zhu2020mnsim,wan2022compute} proposed to directly transfer the partial outputs of a NN layer to the PEs where its next layer is located.
        Such inter-layer parallelism, however, may cause pipeline bubbles \cite{qiao_atomlayer_2018}.

    \subsection{Threat Models and Countermeasures}
\begin{figure}
            \begin{center}
            \includegraphics[width=14cm]{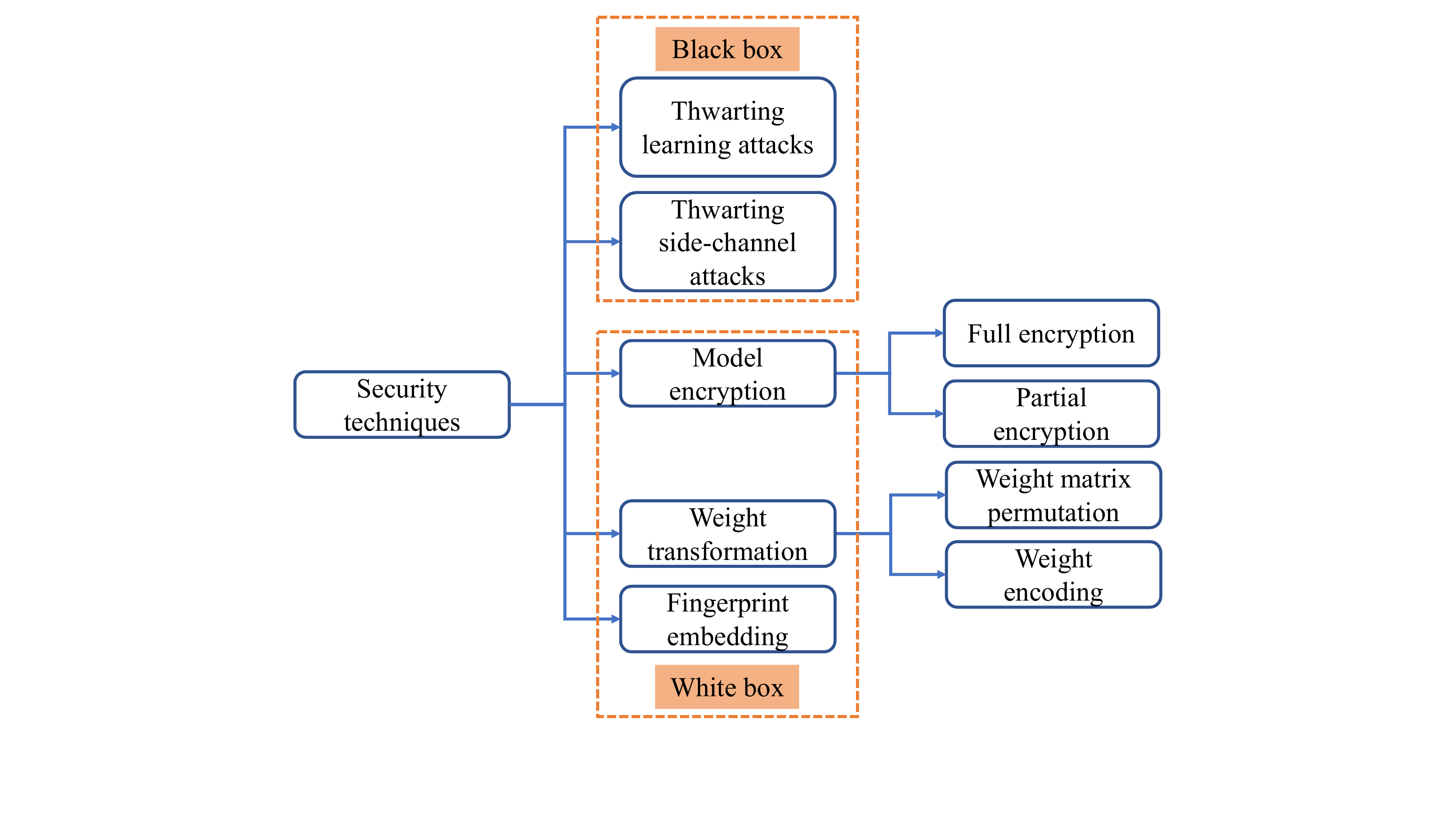}
            \end{center}
            \caption{Security technique categorization: thwarting learning attacks (BB), thwarting side-channel attacks (BB), NN model encryption (WB), NN weight transformation (WB), and fingerprint embedding (WB).} 
            \label{fig:catagorize}
        \end{figure}

        The current commercial memristive computing chips are embedded in boards with M.2 \cite{mythic_mm1076_nodate} or PCIe \cite{mythic_mp10304_nodate} interfaces.
	    The memristive chips may be also equipped with I/Os ports such as GPIOs and I$^2$C \cite{mythic_m1076_nodate}.
        In line with the different security techniques, the adversary with physical access to the boards is assumed to have different capabilities.
        As shown in Fig. \ref{fig:catagorize}, the threat models are divided into BB models and WB models.
        In the BB model, as shown in Fig. \ref{fig:basic_idea}, the adversary can only access the inputs/outputs of the memristor computing systems and cannot access directly the intermediate states of the systems such as the intermediate NN layer outputs or NN weights.
        The adversary could, however, exploit side-channel information such as the memory access patterns or power consumption.
        One common BB model attack is a learning attack, which is based on collecting a certain number of the NN models' input/output pairs \cite{tramer_stealing_2016}.
        Another potential type of attack is a side-channel attack that exploits the systems' covert information \cite{hua2018reverse,batina_csi_2019,dubey2019maskednet,yan2020cache}.
        In the WB model, as shown in Fig. \ref{fig:basic_idea}, the adversary is assumed to be able to read the stored NN weights from the systems.
        This threat model, a result of the nonvolatility of memristor devices, was first widely considered by existing works for the memristor devices’ application as the main memory \cite{young2015deuce,awad2016silent,awad2019triad,zuo2019supermem}.
        For memristor computing systems, the adversary could exploit the universal interfaces of the boards or ports of the chips, and hence the data theft threat is also viable.
        Besides, \cite{huang_new_2020} claimed it is also possible to able to read the stored NN weights from the systems through micro-probing the peripheral circuits of memristor crossbars, which further contributes to the strength of the WB model.
        The existing countermeasures for the WB model manage to prevent the adversary from reading the NN weights correctly.
        They are categorized into three classes: NN model encryption, NN weight transformation, and fingerprint embedding.
        In Sections \ref{sec:bb} and \ref{sec:wb}, we discuss the countermeasures for the BB model and WB model, respectively, in detail.
        

\section{Black-box threats and countermeasures}
    \label{sec:bb}
\begin{figure}[ht]
        \begin{center}
        \includegraphics[width=17cm]{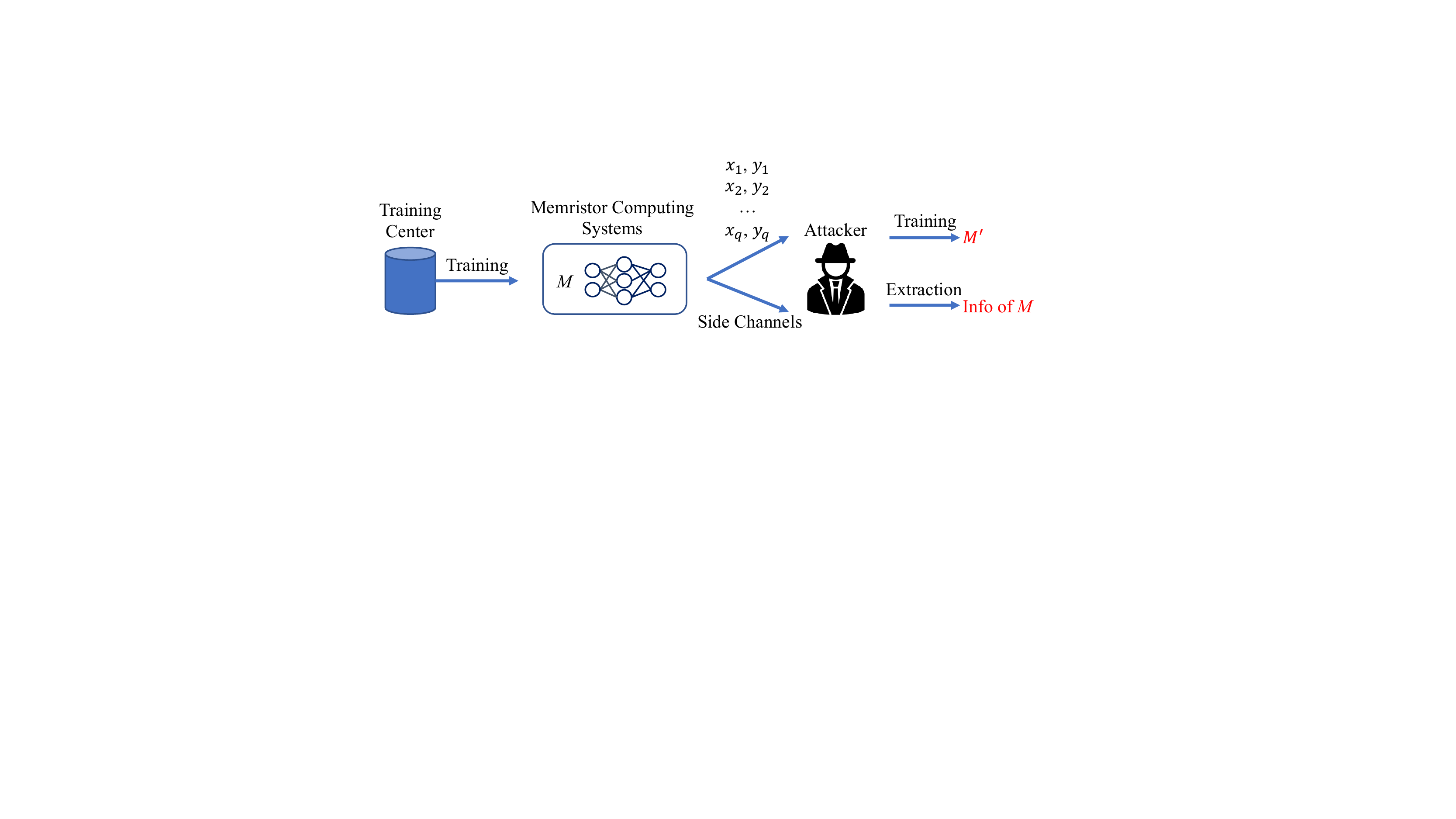}
        \end{center}
        \caption{BB threat model: The attacker cannot access the training dataset but can query the memristor computing system loaded with the well-trained NN \textit{M} in order to train a similar NN $M^{'}$; the attacker can also use side-channel techniques to extract \textit{M}'s confidential information such as its NN structure. }
        \label{fig:bb_model}
    \end{figure}

    For the BB threat model, memristor computing systems are akin to proverbial black boxes as far as concerns the attacker.
    As shown in Fig. \ref{fig:bb_model}, the attacker cannot access the training dataset or the NN weights, but can manipulate the inputs and observe the outputs and exploit side-channel analysis.

    \subsection{Thwarting Learning Attacks}
        Learning attacks are a powerful NN model theft technique.
        The adversary queries the system and observes the outputs.
        Once sufficient input/output pairs are collected, a similar functioning NN model could be trained using the input/output pairs \cite{tramer_stealing_2016}.
        As shown in Fig. \ref{fig:bb_model}, the training center trains a proprietary NN and implements it in an edge device as a service provider, which is accessible to the attacker.
        The attacker queries the devices $q$ times to get $q$ input/output pairs and trains an NN $M^{'}$ that is comparable to the original $M$.

        To thwart this kind of attack, \cite{yang_thwarting_2020} proposed to leverage the obsolescence effect of memristor devices to reduce the inference accuracy of the computing systems for unauthorized users.
        The obsolescence effect of memristor devices causes them to be in the HRS or LRS as a result of read voltage pulses across them.
        The authors of \cite{yang_thwarting_2020} enhanced the obsolescence rate of memristor devices by increasing the voltage amplitude so that the inference accuracy of the computing systems drops dramatically following a limited number of queries.
        To maintain the computing system's inference accuracy, the memristor devices need to be calibrated by writing the original weights on them to counter the obsolescence; however, the authors assumed the calibration process does not work when initiated by unauthorized users \footnote{In their paper, authorized and unauthorized users refer to those who can access the computing systems legally and illegally, respectively}.
        Thus, the attacker, as an unauthorized user, cannot collect sufficient input/output pairs to reverse-engineer the NN models loaded in the computing systems.

        \cite{rajasekharan_scanet_2021} proposed to utilize the innate stochasticity in super-paramagnetic magnetic tunnel junctions (s-MTJs) to thwart learning attacks.
        s-MTJs are unstable and need to be refreshed periodically to keep the weights mapped on them.
        MTJs, different from s-MTJs, are stable and do not need refreshing operations.
        By only storing significant weights on s-MTJs and the other weights on MTJs, the number of cells requiring periodic refreshing is reduced.
        Similar to \cite{yang_thwarting_2020}, the authors in \cite{rajasekharan_scanet_2021} assumed that only authorized users can keep the refreshing operations executing.
        The induced switching in the s-MTJs causes the systems' performance to deteriorate in a very short time if the s-MTJs are not refreshed. Thus, unauthorized users are thwarted in their effort to collect sufficient input/output pairs.
        
    \subsection{Thwarting Side-Channel Attacks}
        Side-channel attacks are another type of non-invasive attack that treat the device under attack as a black box.
        These methods can exploit side channels, such as power, latency, and electromagnetism, to discover confidential information, as shown in Fig. \ref{fig:bb_model}.

        Some works \cite{hua2018reverse,batina_csi_2019,dubey2019maskednet,yan2020cache} have already explored how to steal the NN models through side-channel attacks.
        These attacks either target conventional computing architectures or FPGAs, which are different from memristor computing systems in implementing VMM operations.
        Some of these attacks, however, could be applied to memristor computing systems directly, such as that proposed by \cite{hua2018reverse}, which is focused on memory accessing patterns. 
        \cite{hua2018reverse} assumed that all the intermediate layer outputs are stored in memory and transferred to the computing unit as inputs of the next layer.
        The memory accessing patterns leak the NN structure information.
        For some memristor computing systems, a similar layer-by-layer processing technique is used \cite{qiao_atomlayer_2018,krishnan2021impact}. 
        Thus, the memory accessing patterns could also be a side-channel vulnerability that adversaries can exploit in memristor computing systems.
        To counter this attack, the authors referred to techniques for hiding the memory accessing patterns \cite{goldreich1996software} that depend on oblivious RAM algorithms.

\section{White-box threats and countermeasures}
    \label{sec:wb}
\begin{figure}
        \begin{center}
        \includegraphics[width=14cm]{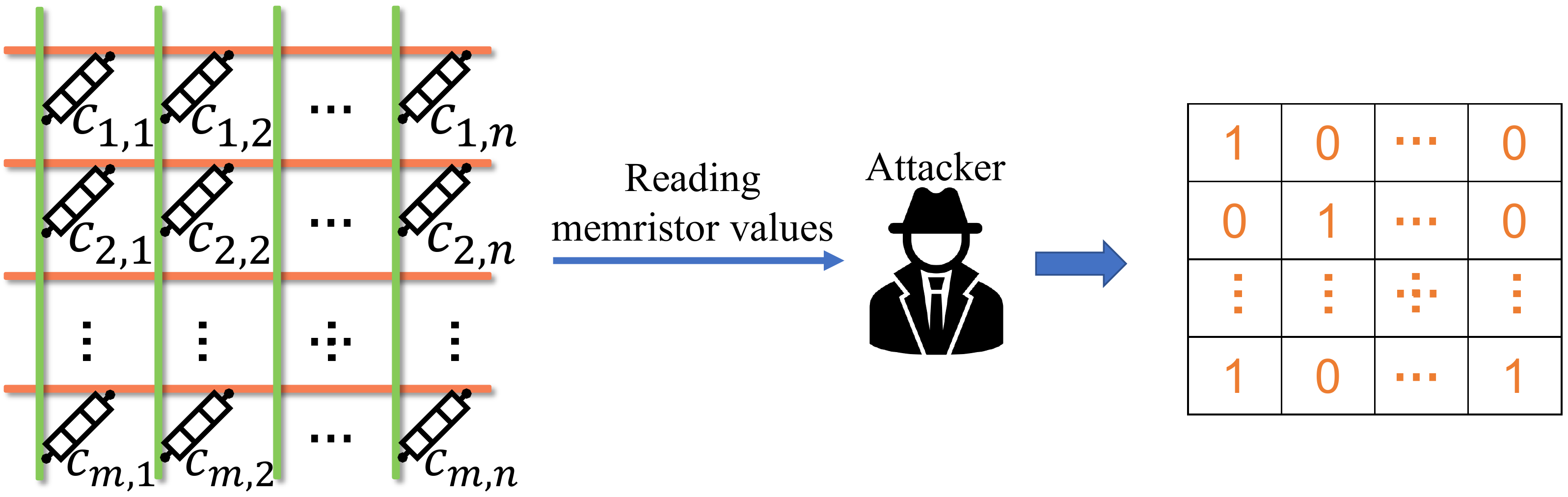}
        \end{center}
        \caption{WB threat model: the attacker can read the memristor values to reverse-engineer the corresponding NN matrices stored in the memristor crossbars.}
        \label{fig:wb_model}
    \end{figure}

    For the white-box threat model, as shown in Fig. \ref{fig:wb_model}, the attacker could read the NN weights stored in the memristor crossbar by exploiting the nonvolatility of memristor devices \cite{young2015deuce,awad2016silent,awad2019triad,zuo2019supermem}. Alternatively, the attacker could use micro-probing techniques \cite{huang_new_2020} so that the NN weight matrices stored in the memristor crossbar can be reverse-engineered.

    \subsection{Model Encryption}
        Methods based on encrypting the NN weights are straightforward.
        Such methods encrypt the NN weights and decrypt them each time they are used.
        The methods require additional write operations to the memristor devices.
        To reduce the latency and high energy consumption produced by the encryption and decryption, \cite{li_p3m_2019,lin_chaotic_2021} focused on accelerating and optimizing these processes and \cite{cai_enabling_2019} focused on reducing the necessary encryption/decryption data.
        One security concern of this type of method is that the NN weights on the memristor crossbars executing the NN algorithms are decrypted and thus exposed to the adversary.
        
        \subsubsection{Full encryption}
            \cite{li_p3m_2019} proposed to encrypt the whole NN model based on XOR before implementing it in the computing systems.
            The encrypted NN model is stored in off-chip memory and copied to the eDRAM buffer and then to the computing units.
            The decryption is done and accelerated in the eDRAM with modifier sense circuits.
            \cite{lin_chaotic_2021} proposed to encrypt the NN weights based on chaotically exchanging the weight positions; however, the encyption/decryption needs additional hardware and is not compatible with memristor crossbars.

        \subsubsection{Partial encryption}
            \cite{cai_enabling_2019} proposed to select the most significant weights of each NN layer.
            The most significant weights are defined as the weight with the largest gradient when their sign bits are flipped.
            The authors showed that only encrypting the MSW can cause the NN models to become useless for inference tasks.
            Compared to the full encryption methods, the partial encryption method reduces the number of encryption/decryption weights significantly (e.g., only 20 per layer for ResNet-101).
        

    \subsection{NN Weight Transformation} 
        In the NN weight transformation protection methods, the protected NN weights are stored in the computing units and there is no need to rewrite the weights so that the energy and latency brought by the weight rewriting operations are eliminated. 
        Thus, this class of methods aims to minimize the imposed hardware overhead.
        Moreover, the weights are always protected -- which is a significant benefit over the model encryption methods.
        
        \subsubsection{Weight matrix permutation}
            \cite{zou_security_2020} proposed to obfuscate the crossbar row connections between positive and negative crossbars.
            During inference, the obfuscating module is configured with the correct keys.
            Without the correct keys, the theft NN model cannot function properly.
            \cite{wang_low_2021} proposed to obfuscate the column connections between consecutive crossbars.
            The crossbars are first divided into smaller groups and then MUX and DEMUX are used to configure the groups of each two consecutive crossbars.
            \cite{huang_new_2020}a \footnote{In \cite{huang_new_2020}, the authors proposed two protection techniques. To distinguish between them, we denote the two techniques as \cite{huang_new_2020}a and \cite{huang_new_2020}b, respectively.} also considered obfuscating the crossbar column connections.
            Their implementation is based on SRAM arrays.
            By configuring the SRAM arrays with one-hot coding for each row, the SRAM arrays could permute the output channels into any order. 
        
        \subsubsection{Weight encoding}
            
            \cite{li2021secure} proposed to encode a weight $w$ as its XORed result $w_e$, which is expressed as
            \begin{equation} \label{eq:xor}
                w_e = w \oplus k,
            \end{equation}
            where $k$ is a key. 
            The memristor crossbars store both $w_e$ and its complementary $\hat w_e$ in the same crossbar column.
            The dot-product between input $i$ and $w$ can be implemented as
            \begin{equation} \label{eq:decoding}
                i \cdot w = i \cdot (w_e \oplus k) = (i \cdot k) \cdot \hat w_e + (i \cdot \hat k) \cdot  w_e,
            \end{equation}
            where $\hat k$ is $k$'s complementary.
            Thus, the decoding for weight $w$ could be done on the input, leaving the VMM operations on the memristor crossbar unaffected.
            Nevertheless, this method requires double the number of memristor devices compared with other methods.
            
            \cite{zou2022} proposed to selectively encode some columns of weights as their 1's complement and leave the other untouched.
            The attacker does not know which columns of weights are encoded so the actual representation of the weights is hidden.
            The authors designed and implemented the protection methods for bias-based mapping and differential mapping, respectively.

    \subsection{Fingerprint Embedding}
        Methods based on fingerprint embedding use the hardware variation of the computing systems as a fingerprint and embed it inside the NN weights.
        The directly stolen NN model cannot function well without the fingerprint.
        The fingerprint is hardware dependent and hard to copy.
        
        \cite{huang_new_2020}b observed that the ADC offset plays an important role in the model accuracy and the NN model needs to be refined to improve the accuracy.
        Thus, the intrinsic ADC offset pattern of a specific chip could be the fingerprint embedded in the NN model.
        The directly extracted NN models do not contain the fingerprint so that the NN models are well-protected.

\section{Cross-comparison}
    We now present a cross-comparison of the limitations of the above security techniques.
    The results are summarized in Table \ref{tab:cross_comparison}.

    \begin{table}
        \centering
        \caption{Cross-comparison of the security techniques}
        \label{tab:cross_comparison}
        \scalebox{0.84}{
            \begin{tabular}{ |c|c|c|c|c|c|c| } 
                \hline
                                                  &\begin{tabular}{@{}c@{}}Stop authorized \\ adversary?\end{tabular}
                                                  &\begin{tabular}{@{}c@{}}Structures (A) or \\ parameters (B)?\end{tabular}
                                                  &\begin{tabular}{@{}c@{}}Require \\ retraining?\end{tabular}
                                                  &\begin{tabular}{@{}c@{}}Memristor \\ writing?\end{tabular}
                                                  &\begin{tabular}{@{}c@{}}Analog (A) or \\ digital (B)?\end{tabular}
                                                  &\begin{tabular}{@{}c@{}}Full-time \\ protection?\end{tabular}
                                                  \\ \hline
                \cite{yang_thwarting_2020}        &No  &B               &No  &Yes &A        &N/A   \\ \hline
                \cite{rajasekharan_scanet_2021}   &No  &B               &No  &Yes &A and B  &N/A   \\ \hline
                \cite{hua2018reverse}             &Yes &A               &No  &No  &A and B  &N/A   \\ \hline
                \cite{li_p3m_2019}                &Yes &B               &No  &Yes &A and B  &No    \\ \hline
                \cite{lin_chaotic_2021}           &Yes &B               &No  &Yes &A and B  &No    \\ \hline
                \cite{cai_enabling_2019}          &Yes &B               &No  &Yes &A and B  &No    \\ \hline
                \cite{zou_security_2020}          &Yes &B and partial A &No  &No  &A and B  &Yes   \\ \hline
                \cite{wang_low_2021}              &Yes &B and partial A &No  &No  &A and B  &Yes   \\ \hline
                \cite{huang_new_2020}a            &Yes &B and partial A &No  &No  &B        &Yes   \\ \hline
                \cite{li2021secure}               &Yes &B               &No  &No  &B        &Yes   \\ \hline
                \cite{zou2022}                    &Yes &B and partial A &No  &No  &A and B  &Yes   \\ \hline
                \cite{huang_new_2020}b            &Yes &B               &Yes &No  &A and B  &Yes   \\ \hline
            \end{tabular}
        }
    \end{table}

    \begin{itemize}
        \item \textbf{\textit{Ability to protect from attacks by an authorized adversary}}: 
        The adversary could be an authorized or an unauthorized user.
        As an authorized user, the adversary could carry out an unlimited number of inference tasks on a memristor computing system.
        In this scenario, the countermeasure of thwarting the learning attacks \cite{yang_thwarting_2020} \cite{rajasekharan_scanet_2021} would not stop the adversary.
        In the other countermeasures, the keys of the protection methods can be configured as internal and not accessible by any users; thus they would work regardless of whether or not the adversary is authorized.  
        
        \item \textbf{\textit{NN structures and NN parameters}}: 
        The secrecy of a NN model includes both its structure and its parameters.
        For memristor computing systems, the countermeasure referred by \cite{hua2018reverse} protect only the NN structures.
        Some NN weight transformation methods \cite{zou_security_2020,wang_low_2021,huang_new_2020,zou2022} exploit the padded fake rows or columns to hide the exact size of weight matrices with fewer rows or columns than the size of the crossbar.
        As a result, the structures of these NN layers with small weight matrices are protected.
        Other countermeasures protect only the NN parameters.
        Note that learning attacks would not succeed easily without knowledge of the NN structures \cite{batina_csi_2019}. 
        The learning attacks thus could be treated as a way to steal the NN parameters.
        Hence, the methods proposed by \cite{yang_thwarting_2020} and \cite{rajasekharan_scanet_2021} are deemed to protect the NN parameters only.
        
        \item \textbf{\textit{Requiring retraining?}} 
        \cite{huang_new_2020}b requires the NN parameters to be retrained so that they are compatible with the hardware variation.
        This kind of method requires information about the target devices in advance of mapping the NN models to them.
        Retraining the NN models for each specific hardware systems, however, may be time-consuming and not practical.
        In the other countermeasures, retraining NN parameters is not necessary.
        
        \item \textbf{\textit{Additional memristor write operations:}} 
        Both full \cite{li_p3m_2019,lin_chaotic_2021} and partial \cite{cai_enabling_2019} encryption techniques require additional write operations to the memristor devices for each inference.
        In the techniques suggested by \cite{yang_thwarting_2020} and \cite{rajasekharan_scanet_2021} the memristor devices must be periodically written because of the devices' resistance drifting.
        Additional memristor write operations not only consume high energy and introduce long latency into the systems~\cite{chang_194_2014, yao_face_2017}, they also shorten the lifetime of the memristor computing systems due to the limited endurance of memristor devices~\cite{wang_cross-point_2019}
        The other countermeasures do not require additional memristor write operations.
        
        \item \textbf{\textit{Analog or digital memristor devices}}: 
        In \cite{yang_thwarting_2020}, the authors only investigated the obsolescence effect in analog memristor devices.
        Determining whether this technique works for digital memristor devices needs further justification since the resistance states of digital memristor devices have larger margins than analog devices.
        \cite{huang_new_2020}a only works for digital devices because the SRAM arrays can only output binary results at the BLs.
        The technique proposed by \cite{li2021secure} also only works for digital devices because the XOR-based encoding is only applicable for single-bit-precision devices.
        The other countermeasures would work for both kinds of devices. 
        
        \item \textbf{\textit{Full-time protection}}: 
        Full-time protection means that the systems are always protected whenever the adversary carries out an attack.
        Among the countermeasures of the WB model, the model encryption methods \cite{li_p3m_2019,lin_chaotic_2021,cai_enabling_2019} decipher the NN weights and store them as plaintext when they are involved in NN computing. 
        The adversary may analyze the systems’ side channels, such as power consumption, to pinpoint when the NN weights are involved in computing. 
        Once the specific point in time is determined, the adversary could turn off the system and then read the unprotected NN weights from non-volatile memristors.
        Though \cite{cai_enabling_2019} narrowed the attack window by exposing only one NN layer at a time to the adversary, the threat is not thwarted. Conversely, the methods based on NN weight transformation leave no such attack window and protect the NN weights all the time.
    \end{itemize}

\section{Discussion}
    We now discuss the developing directions of security techniques for protecting NN models in memristor computing systems.
    
    \begin{itemize}
        \item \textbf{\textit{Countermeasures against comprehensive attacks}}:
            Each reviewed countermeasure can only prevent one type of specific attack.
            Sophisticated attackers, however, may maximize their attack strategies by combining a number of different attacks.
            For example, the adversary may use side-channel attacks to infer the NN structures and micro-probing techniques to extract the NN weights.
            Simply adding more and more defense techniques may impose a heavy hardware overhead and even open up new opportunities for the adversary to exploit.
            Thus, more thoughtful countermeasures are required both to deliver protection for multiple attacks and ensure low hardware overhead.  
            
        \item \textbf{\textit{More investigations into side-channels attacks}}:
            Side-channel attacks against other NN computing systems, such CPUs and FPGAs \cite{yan2020cache,dubey2019maskednet}, have attracted much attention. 
            To date, we have seen few The side-channel attacks against memristor computing systems.
            Though different from other NN computing systems, memristor computing systems have a lot in common with them such as activation functions, pooling functions, and storing intermediate results in buffers.
            Side-channel attacks focusing on those attack surfaces in other NN computing systems may also be feasible in memristor computing systems.
            Thus, more investigations into side-channel attacks are needed.
            
        \item \textbf{\textit{Alternative approaches to thwarting learning attacks}}:
            The current learning attack countermeasures can only prevent attacks by unauthorized users.
            Authorized users, however, are neither barred nor limited in their access to the input/output pairs.
            Limiting authorized users' access to input/output pairs would impose an inconvenience for them, making this avenue an unrealistic option.
            Alternative approaches to thwarting learning attacks must consider learning attacks by authorized users.
            We suggest looking for alternate solutions from the perspectives of NN model training strategies or training data manifestation.
    \end{itemize}
    
\section{Conclusions}
    In recent years, memristor computing systems have dramatically improved the energy efficiency of NN algorithms. 
    The security issues of memristor computing systems, however, should be addressed before they become widely used commercially. 
    Specifically, the NN models loaded in memristor computing systems face potential theft threats because of memristors’ nonvolatile nature. 
    In this paper, we reviewed the existing security techniques for protecting NN models for memristor computing systems. 
    Two threat models assume the adversary has different attack capabilities, i.e., the BB model and the WB model. 
    For the threat models, based on the attack method, we classified the existing countermeasures into five sub-classes. 
    Those countermeasures are limited because of their defense capabilities and hardware implementation. 
    For example, the countermeasures against learning attacks can only protect the memristor computing systems from unauthorized users; the protection methods based on NN model encryption leave an attack window open when NN weights are involved in computing. 
    We presented a cross-comparison of the limitations of the existing security techniques and suggested future research directions for this field.
    
\section*{Author Contributions}
    Minhui Zou: Proposing the methodology, investigating the related works, designing the security metrics, and writing the original manuscript.
    Nan Du: Proposing the methodology and reviewing and editing the manuscript.
    Shahar Kvatinsky: Reviewing and editing the manuscript and supervising this project.

\section*{Funding}
    This paper acknowledges the funding by the German Research Foundation (DFG) Projects MemDPU (Grant Nr. DU1896/3-1), MemCrypto (Grant Nr. DU 1896/2-1), and the European Union's Horizon 2020 Research And Innovation Programme FET-Open NEU-Chip (Grant agreement No. 964877). 



\bibliographystyle{styles/Frontiers-Harvard} 
\bibliography{references}

\end{document}